# Experiments on bright field and dark field high energy electron imaging with thick target material

**Authors:**
Zheng Zhou,[1] Yingchao Du,[1] Shuchun Cao,[2] Zimin Zhang,[2,*] Wenhui Huang,[1,#] Huaibi Chen,[1] Rui Cheng,[2] Zhijun Chi,[1] Ming Liu,[2] Xiaolu Su,[1] Chuanxiang Tang,[1] Qili Tian,[1] Wei Wang,[1] Yanru Wang,[2,4] Jiahao Xiao,[2,4] Lixin Yan,[1] Quantang Zhao,[2] Yunliang Zhu,[2,4] Youwei Zhou,[2,4] Yang Zong,[2] and Wei Gai [1,3]

[1] Department of Engineering Physics, Tsinghua University, Beijing 100084, China

[2] Institute of Modern Physics, Chinese Academy of Sciences, Lanzhou 730000, China

[3] Argonne National Laboratory, Lemont, Illinois 60439, USA

[4] University of Chinese Academy of Science, Beijing 100049, China

**Abstract:**

Using a high energy electron beam for the imaging of high density matter with both high spatial-temporal and areal density resolution under extreme states of temperature and pressure is one of the critical challenges in high energy density physics (HEDP). When a charged particle beam passes through an opaque target, the beam will be scattered with a distribution that depends on the thickness of the material. By collecting the scattered beam either near or off axis, so-called bright field or dark field images can be obtained. Here we report on an electron radiography experiment using 45MeV electrons from an S-band photo-injector, where scattered electrons, after interacting with a sample, are collected and imaged by a quadrupole imaging system. We achieved a few micrometers (~ 4μm) spatial resolution and about ~ 10μm thickness resolution for a silicon target of 300–600 micron thickness. With addition of dark field images that are captured by selecting electrons with large scattering angle, we show that more useful information in determining external details such as outlines, boundaries and defects can be obtained.

**Introduction:**

High Energy Density Physics (HEDP) is the study of matter properties under extreme pressure and temperature conditions, such as those occurring in heavy ion or laser drive fusion and similar processes. Charged particle beam-based diagnostics have been the subject of studies for many years, beginning with the pioneering efforts of F. Merrill et al. [1-5]. In the last decades, there has been strong interest in using electron beams to probe the dynamic processes in a target when driven by either laser or heavy ion beams, and in addition, as an online diagnostic for heavy ion induced target damage [6-8]. The process uses a magnetic imaging lens system (quadrupoles or solenoids) focusing a beam passing through a thick target. The beam is collimated at a focal point, then projected on a luminescent screen [9]. This method is fundamentally different from typical shadowgraphs with either charged particles or X/gamma rays. Several experiments have been performed with results in a good agreement with predications [10-12]. In general, those experiments required a dynamic process with a temporal resolution of ns, and a spatial resolution of sub micrometers. In addition, the diagnostics system must have a large dynamic range and be sensitive to a mixture of high and low Z (atomic number) components in order to understand the hydrodynamics of the HED sample [8]. In the last few decades, many facilities using RF photocathode techniques enabled generation of a pulse train of electrons with high charges (> nC) and high energies (~ 100 MeV), such as at Tsinghua Thomson scattering X-ray source (TTX) [13]. Recently, an experiment using a compact permanent magnet quadrupole based lens was carried out at UCLA, where the micron-size features of a metal sample were imaged with single-shot, low energy (4MeV) electron bunches [14].

In this paper, we report on a new experiment with picosecond electron pulses at Tsinghua University TTX, in collaboration with the Institute of Modern Physics of China. The experimental layout [15] is shown in Figure 1. The electron beam is produced and accelerated to about 45MeV with an S-band photoinjector and accelerator. The average bunch charge is 200 pC and the bunch duration is 10ps. Then the electron beam is transported to the target area after a 90-degree bend. After the e-beam passes through the target, scattered electrons travel through focusing quadrupole arrays which act in the same way as a light optics imaging system. An aperture in both the x and y planes is located at the focal plane or the so-called Fourier plane, 1.3m downstream of the object plane, where the positions of particles are only related to their initial spread angle. Electrons selected by the aperture hit the imaging plane located 2.6m downstream, and the beam patterns are captured by a CCD detector. The main target used in this experiment is an opaque 300μm thick silicon slab with lines engraved with different patterns and depths. A standard 200 mesh molybdenum TEM grid whose hole width is 90μm and bar width is 35μm is used to determine the spatial resolution of imaging system. A radiograph of this grid is shown in Fig 2. By studying the step in transmission across the edge of a bar, the edge spread function can be determined, as described in ref 10.

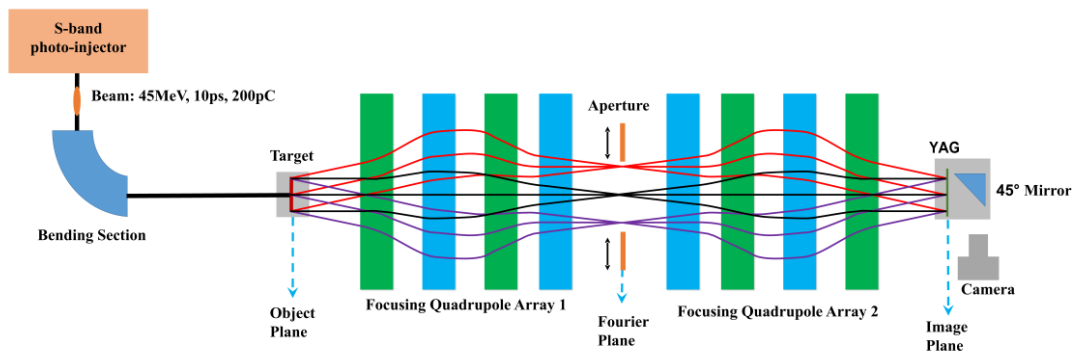

**Fig. 1 experimental setup.** The aperture in the vertical or the horizontal plane comprises two plates which can move independently over a wide range of distances, selecting on-axis or off-axis particles to form a bright field or dark field image.

By fitting the RMS width of the line spread and known geometry of the mesh target, we have determined that the spatial resolution on the object plane is 4 μm, which is very close to the ultimate resolution limited by the 30 micron thick YAG screen and the detection system as shown in Figure 2.

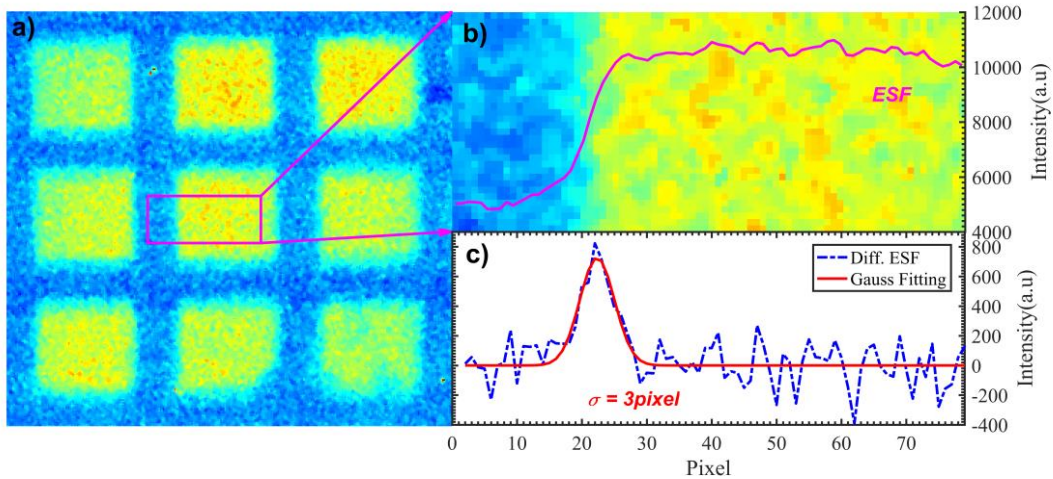

**Fig 2. TEM grid image for spatial resolution calibration.** a) The 200 mesh grid image. b) Zoom to a small region between the bar and the hole (pink rectangular box) and calculate the intensity projection on the x axis, i.e. the edge spread function (pink curve). c) line spread function and its Gaussian fit, giving a resolution of 3 pixels.

**Thick target imaging**

A set of thick targets with etched grooves, which have a known width and depth, are used to study both the spatial and areal density distribution, as shown in Figure 6a. A logo with the etched letters "HEER" (75 micron width) was measured experimentally, revealing qualitative surface profile information, as shown in Fig 3. For comparison, we have also acquired images of the sample using a scanning electron microscope, and measured the depth at several points using a high resolution profiler.

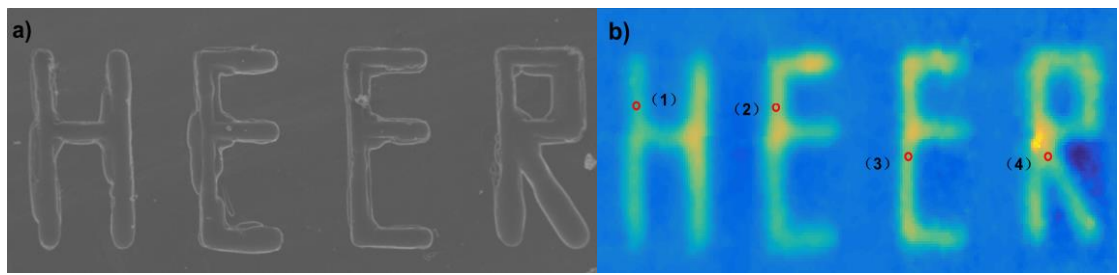

**Fig 3. HEER logo SEM image and radiography image comparison.** a) the profile of the target as scanned by SEM. Some imperfections on the surface and grooves can be clearly observed. b) False color radiographic map of HEER logo image. The depth of the four spots in (b) are (rom left to right) measured to be 42μm, 50μm, 46μm and 54μm respectively.

The letters are clearly visible in the radiographs, but far more detailed features of the logo can also be seen. For example, the non-uniformity of the width in the junctions of the perpendicular grooves seen from the radiographs matches well with the SEM surface profile measurement, especially in the cases of the first letter "E" and the letter "R". In addition, we have also measured four different typical points of the logo as shown in Fig 3(b). The relation between the groove depth and the intensity is rather simple correlates well with the drop in transmission rate as the target thickness increases.

To quantitatively study this system, we have concentrated on small area of a different target etched with the letter "N". The surface profile of this letter is obtained by SEM, as shown in Fig 4(a). The beam intensity in (Fig 4(b)) shows a very nonuniform distribution. The path along which the depth measurement was made is indicated by the dashed red line. When comparing the depth and intensity distribution, we find that a depth difference of a few tens of microns can be resolved, which is equivalent to a density resolution of a few percent.

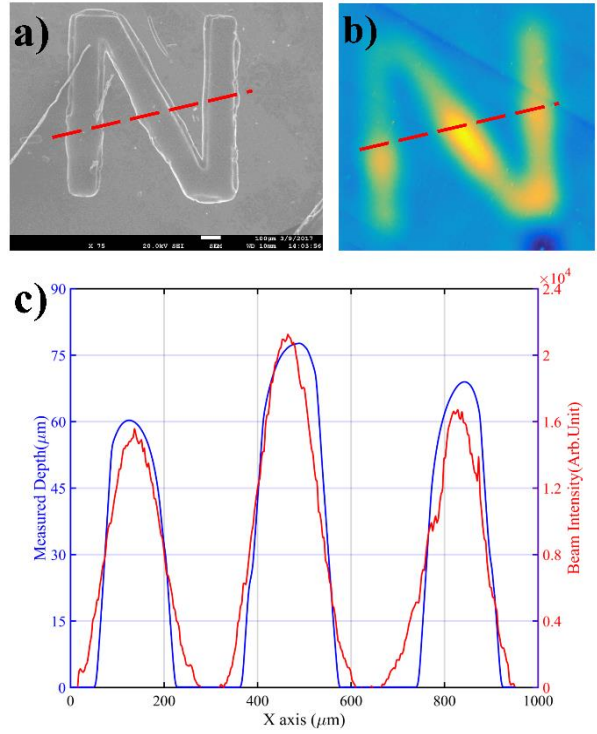

**Fig 4. Letter N image.** a) Letter N surface profile under x75 zoom SEM. The roughness and non-uniformity of grooves due to the etching process is apparent. b) Beam intensity distribution radiograph. The signal of beams passing through the entire target area is subtracted as background, and regions with brighter colors represent the grooves. c) Depth information required by a high resolution step profiler (blue line) along the red dashed path in (b), compared to the intensity distribution along the same path.

**Dark field imaging**

Next, we attempted to image the target using the large angle scattered beam electrons, common referred to as dark-field imaging, in contrast to the on-axis bright field imaging. This is analogous to the contrast enhancing techniques developed for light-optical systems, conventional transmission electron microscopes as well as by the X-ray imaging community [16]. However, dark field imaging, or more basically scattering-based imaging, using high energy charged particles and exhibiting a good signal-to-noise ratio, have not yet been observed previously as far as we know.

In general, the MCS angle is related to beam energy and target thickness as in formula (1),

$$\frac{N(l,\theta_c)}{N_0} = \frac{1}{\theta_0(l)\sqrt{2\pi}} e^{-\frac{1}{2}\left(\frac{\theta_c}{\theta_0(l)}\right)^2} \quad (1)$$

with $\theta_c$ is the collection angle of the aperture. For a quasi mono-energetic electron beam, the Gaussian distribution width is related to the target thickness, with a thicker target having a wider FWHM of the angular distribution, as in the case indicated by the blue line in Fig 5(a).

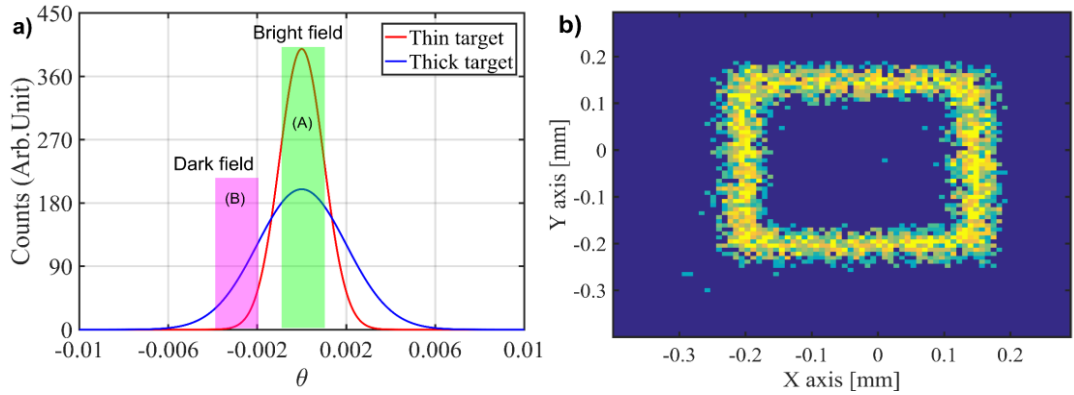

**Fig 5. Dark field imaging simulation.** a) Principle of dark field imaging. A probe beam passing through a thicker target will form a wider Gaussian angular distribution. Selecting electrons with larger scattered angle will form a dark field image (pink region). b) Simulation result using a sample holder with a hole in the center. The bright square region represents the margin of the hole.

In our study, a 2mm thick sample holder made of stainless steel with a 3mm square hole in the center is used for the dark field imaging simulation. In the bright field situation, electrons pass through the hole without much energy loss or angular spread will form the bright field image. In the dark field situation, electrons striking the hole edge will suffer considerable energy loss and angular spread, and these electrons will be selected to pass through the off-axis aperture.

With stepper motors to control the focus aperture location and dimensions, with the aperture placed on the axis center, as shown in area (A) in Fig 5(a), the ratio of slightly scattered electrons or 'unscattered electrons' to the overall transmitted electrons is much larger than the scattered electrons, which is the normal bright-field image case. While placed off axis, as shown in area (B) in Fig 5(a), only scattered electrons are allowed to pass, and result in a dark-field image. In reality, transition from bright-field image to dark field image can be observed by carefully moving the aperture off the axis, while the whole brightness of the image drops dramatically with the decrease of transmitted electrons. The final results are shown in Fig 6. Comparing to the physical features of the sample, it shows good agreement with the patterns of electron images. where the difference in thickness between the hollow part and the edge creates the contrast, and the roughness of the target edge can be observed. Especially in the case of the region in the red circle of picture (b), where the intensity nonuniformity of the right border corresponds to the CCD measurement result. This nonuniformity is proved to be caused by the scratch on the right border, as shown in picture (c).

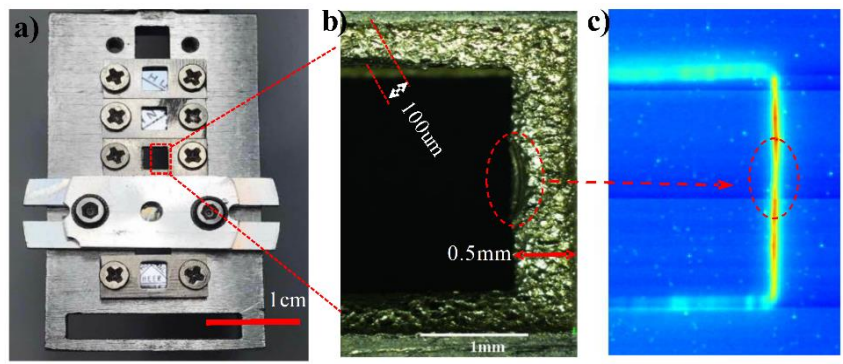

**Fig 6. Installation, zoom view and dark field images of hollowed target.** a) Photograph of targets and target holder used in the experiment. b) Zoom of hollow target in a microscope image. c) Dark field image of the target.

In addition, we have also observed a dark field image from the mesh target, with one grid square that was filled with glue (the lighter spot in the left bottom of the grid in Fig 6a) which can be observed as either by on axis bright field (Fig 6b) or dark field imaging (Fig 6c). This shows how dark-field imaging can be used for defect identification in the target.

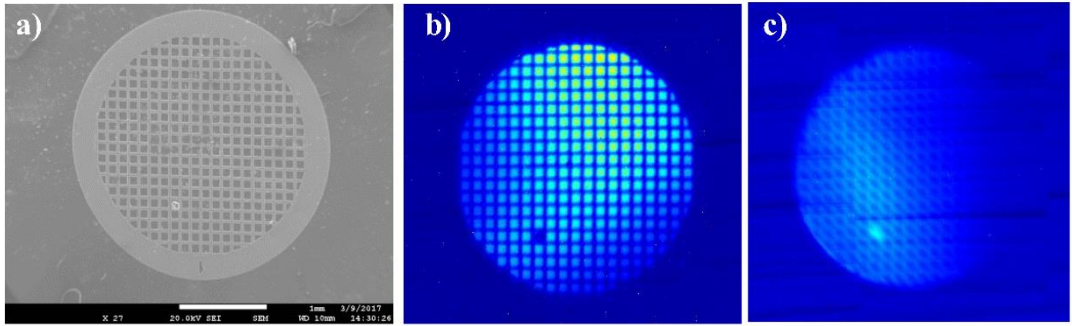

**Fig 7. SEM, bright field and dark field images of TEM grid.** a) The SEM image of the 200 mesh standard TEM grid. b) The bright field image. c) The dark field image.

In summary, we have performed high energy electron radiography on an opaque target with a spatial resolution of 4 micrometers and areal density resolution to 10μm accuracy. Further resolution improvement can be achieved by optimizing the beam transport, particularly the magnification. By further magnifying by a factor of 10, submicron resolution can easily be achieved. This technique is a promising candidate for HEDP or Inertial Confinement Fusion (ICF) diagnostics. We have also observed dark-field imaging with the off-axis aperture technique for the first time in electron radiograph experiments, thus extra and complementary information of the target can be obtained in addition to the on-axis bright-field imaging.

Furthermore, with the expected picosecond time resolution via the picosecond or even femtosecond electron bunches from the existing photocathode RF gun the electron imaging system can be used in other dynamic systems, as long as there is a density distribution. Even multiple axis imaging of the target can yield 3-D images [17]. If a target is much denser or thicker, then one can increase the electron beam energy correspondingly, with the physics otherwise the same as in this experiment.

zzm@impcas.ac.cn
huangwh@mail.tsinghua.edu.cn

This work is supported by the National Natural Science Foundation of China (NSFC Grants No. 11435015) and the Program of International S&T Cooperation by the Ministry of science and technology of China (Grants No. 2016YFE0104900).

**References**

[1]  N. S. P. King *et al.*, Nucl. Instrum. Methods Phys. Res., Sect. A 424, 84 (1999).
[2]  S. A. Kolesnikov et al., High Pressure Research 30, 83 (2010).
[3]  Y. M. Antipov et al., Instruments and Experimental Techniques 53, 319 (2010).
[4]  P. A. Rigg et al., Physical Review B 77 (2008).
[5]  C. L. Morris, N. S. King, K. Kwiatkowski, F. G. Mariam, F. E. Merrill, and A. Saunders, Reports on progress in physics. Physical Society 76, 046301 (2013).

[6] Merrill F E, "Charged Particle Radiography for MaRIE", LA-UR-16-26776, 2016.

[7] Y. T. Zhao, Z. M. Zhang, H. S. Xu, W. L. Zhan, W. Gai, J. Q. Qiu, S. C. Cao, C. X. Tang, A HIGH RESOLUTION SPATIAL-TEMPORAL IMAGING DIAGNOSTIC FOR HIGH ENERGY DENSITY PHYSICS EXPERIMENTS Proceedings of IPAC2014, Dresden, Germany, THOAB03

[8] Wei Gai, Jiaqi Qiu, Chunuang Jing. Electron imaging system for untrafast diagnostics of HEDLP. Proc. of SPIE Vol. 9211, 921104.

[9] C. T. Mottershead and J. D. Zumbro, in Proceedings of the 1997 Particle Accelerator Conference (IEEE, Vancouver,1997), Vol. 1, pp. 1397-1399.

[10] F. Merrill, F. Harmon, A. Hunt, F. Mariam, K. Morley, C. Morris, A. Saunders, and C. Schwartz, Nuclear Instruments and Methods in Physics Research Section B: Beam Interactions with Materials and Atoms 261, 382.

[11] F. E. Merrill, C. L. Morris, K. Folkman, F. Harmon, A. Hunt, and B. King, in Proceedings of the 2005 Particle Accelerator Conference (IEEE, Knoxville, TN, 2005), pp. 1715-1717.

[12] Zhao, Quantang, et al. "High energy electron radiography experiment research based on picosencond pulse width bunch." *Proc. of LINAC2014* (2015): 4.

[13] C. Tang et al., Nuclear Instruments and Methods in Physics Research Section A: Accelerators, Spectrometers, Detectors and Associated Equipment 608, S70 (2009).

[14] Cesar et al., Phys Rev Lett 117, 024801 (2016).

[15] Q. Zhao et al., Nuclear Instruments and Methods in Physics Research Section A: Accelerators, Spectrometers, Detectors and Associated Equipment 832, 144 (2016).

[16] F. Pfeiffer, M. Bech, O. Bunk, P. Kraft, E. F. Eikenberry, C. Bronnimann, C. Grunzweig, and C. David, Nat Mater 7, 134 (2008).

[17] Y. Zhao et al., Laser and Particle Beams 34, 338 (2016).